# Ion Transport through Short Nanopores Modulated by Charged Exterior Surfaces

Long Ma,[1,2] Zhe Liu,[1] Bowen Ai,[1] Jia Man,[1] Jianyong Li,[1] Kechen Wu,[2] and Yinghua Qiu[1,2,3]*

1. Key Laboratory of High Efficiency and Clean Mechanical Manufacture of Ministry of Education, National Demonstration Center for Experimental Mechanical Engineering Education, School of Mechanical Engineering, Shandong University, Jinan, 250061, China

2. Fujian Key Laboratory of Functional Marine Sensing Materials, Minjiang University, Fuzhou 350108, China

3. Shenzhen Research Institute of Shandong University, Shenzhen, Guangdong, 518000, China

*Corresponding author: yinghua.qiu@sdu.edu.cn

# ABSTRACT

Short nanopores find extensive applications capitalizing on their high throughput and detection resolution. Ionic behaviors through long nanopores are mainly determined by charged inner-pore walls. When pore lengths decrease to sub-200 nm, charged exterior surfaces provide considerable modulation to ion current. We find that the charge status of inner-pore walls affects the modulation of ion current from charged exterior surfaces. For 50-nm-long nanopores with neutral inner-pore walls, charged exterior surfaces on the voltage (surface$_V$) and ground (surface$_G$) sides enhance and inhibit ion transport by forming ion enrichment and depletion zones inside nanopores, respectively. For nanopores with both charged inner-pore and exterior surfaces, continuous electric double layers enhance ion transport through nanopores significantly. The charged surface$_V$ results in higher ion current by simultaneously weakening ion depletion at pore entrances and enhancing the intra-pore ion enrichment. The charged surface$_G$ expedites the exit of ions from nanopores, resulting in a decrease in ion enrichment at pore exits. Through adjustment in the width of charged-ring regions near pore boundaries, the effective charged width of the charged exterior is explored at ~20nm. Our results may provide a theoretical guide for further optimizing the performance of nanopore-based applications, like seawater desalination, biosensing, and osmotic energy conversion.

## Keywords:

Nanopores, Electric double layers, Ion concentration polarization, Ion enrichment/depletion

## I. INTRODUCTION

Nanopores act as important channels for the transport of ions and fluids in various fields, such as biosensing,[1] ionic circuits,[2] desalination,[3] and energy conversion.[4] The development of nanofabrication has enabled the successful creation of nanopores with diameters ranging from several micrometers to sub-1 nanometers.[5, 6] Under such highly confined spaces, charged inner-pore surfaces govern the transport processes of charged species, such as ions and biomolecules, through nanopores due to the long-range electrostatic interactions.[7] Also, surface charges on inner-pore walls enable the nanopore with ionic selectivity[8-10] to counterions due to the formation of electric double layers (EDLs), which may induce ion concentration polarization (ICP).[11] These phenomena are particularly obvious in pores with a large length-diameter ratio, such as polymer nanopores with several micrometers in length.

In earlier simulations[12-16] and theoretical analyses,[8, 17, 18] the surface charges on inner-pore walls are the key points to modulate the ion current through nanopores. When the nanopore's inner surface is locally charged to form a unipolar structure, the asymmetric charge distribution on the inner surface significantly modulates the ion transport through the nanopore.[19-21] Under different polarity voltages, obvious ion enrichment and depletion occur at the junction of charged and uncharged interfaces. The electric potential changes smoothly along the nanopore as the ion enrichment forms in the pore, which creates a small electric field strength. Conversely, due to the large local resistivity in the depletion region, the electric potential in the junction region drops sharply, and the induced large electric field strength severely hinders ion's passage through the pore.[22]

Recent progress in nanofluidic research shows that charged exterior surfaces can play important roles in regulating the ionic flux and ionic selectivity of nanopores[23-26]. Hsu et al.[27] constructed nanopores with only the inner surface modified by a polyethylene (PE) layer and all nanopore surfaces modified by PE, respectively. They found that charged exterior surfaces can reduce ion concentration polarization and obtain a larger ion current. With experiments, through controlled modification to nanopore surfaces with DNA-based functional elements, Xia et al.[28] explored the ion-gating behaviors influenced by different modified surfaces of long nanochannels. They found that functional elements on outer pore walls can produce a synergistic ion-gating with functional elements on inner-pore walls. Létant et al.[29] used a focused ion beam (FIB) to fabricate a single silicon nitride pore with a length of 700 nm and locally derivatized the pore entrances by controlling the growth of oxide rings, followed by preferential functionalization of the silicon oxide with silane chemistry. They found that the current intensity increases as the length of the functionalized exterior surface increases. With short bipolar-charged nanopores, surface charges on exterior surfaces determine the ion current behaviors.[23, 30] When the nanopore length is reduced to the nanoscale, i.e. less than 100 nm, the charged exterior surfaces can effectively promote the diffusion and migration of counterions through nanopores.[26, 31, 32] Also, with the modulation of the ion concentration inside nanopores by charged membranes, significant ion current rectification is induced in a 100-nm-long conical nanopore by the individual charged exterior surface on the tip side.[24]

In this study, we systematically studied the modulation mechanism of ion transport by charged exterior surfaces. For nanopores with a length of 200 nm and above, charged inner surfaces play a dominant role in governing ion transport. While in nanopores shorter than 50 nm, the charged exterior surfaces predominantly determine the ion transport. The streamlines of fluid flow reveal that counterions have distinct migration in the EDL regions near the charged exterior surfaces. The exterior surfaces on the voltage and ground sides can enhance and suppress the ion current by forming ion enrichment and depletion at the entrance and exit of the nanopore, respectively. The phenomena are correlated with the charged status of the inner surface. With the combination of the charged inner-pore surface and the exterior surface, EDLs near charged pore walls provide a continuous passageway for the fast transport of counterions, which enhances the ion concentration inside the nanopore. The high selectivity of nanopores to counterions leads to significant ion depletion and enrichment at the entrance and exit of the nanopore, which is different from the cases with neutral inner-pore surfaces. In the cases with the neutral and charged inner-pore walls, the ion current is dominated by the exterior surface on the ground side and the voltage side, respectively. Our results clarify the ionic behaviors modulated by the charged exterior surfaces, which may shed light on the wide application of short nanopores in related fields.

## II. SIMULATION DETAILS

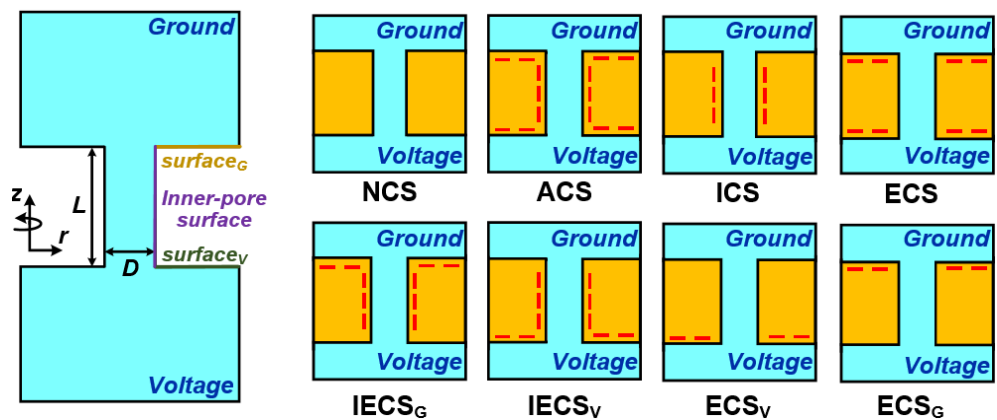

Figure 1 Schematic illustration of simulations. *L* and *D* represent the length and diameter of nanopores. Negatively charged surfaces are shown with red dashed lines.

COMSOL Multiphysics was used to simulate the ion and fluid flow. As shown in Figure 1, cylindrical reservoirs with a fixed radius and height of 5 μm are connected by cylindrical nanopores, which form a sandwich-like structure. The diameter was set to 10 nm. The pore length varied from 10 to 1000 nm, where 50 nm was the default value. KCl solution was used with 100 mM in concentration, which is usually applied in biosensing experiments because $K^+$ ions share a similar diffusion coefficient to $Cl^-$ ions, which are $1.96\times10^{-9}$ and $2.03\times10^{-9}$ $m^2/s$, respectively. A voltage ranging from 0 to 1 V was applied across the nanopore. Since counterions are the main current carriers in charged nanopores, the entrance and exit of the nanopore are defined as the positions where counterions enter and exit the nanopore. The temperature and dielectric constant of water were set to 298 K and 80. Poisson-Nernst-Planck and Navier-Stokes equations were used to describe the electrostatic potential distribution, the ion distribution near charged surfaces, and fluid flow in the system, equations 1-4.[9, 33]

$$\varepsilon \nabla^2 \varphi = -\sum_{i=1}^{N} z_i F C_i \quad (1)$$

$$\nabla \cdot \mathbf{J}_i = \nabla \cdot \left( C_i \mathbf{u} - D_i \nabla C_i - \frac{F z_i C_i D_i}{RT} \nabla \varphi \right) = 0 \quad (2)$$

$$\mu \nabla^2 \mathbf{u} - \nabla p - \sum_{i=1}^{N} (z_i F C_i) \nabla \varphi = 0 \quad (3)$$

$$\nabla \cdot \mathbf{u} = 0 \quad (4)$$

in which $\varphi$ represents the electrical potential. $N$ and $\varepsilon$ denote the number of ion species and the dielectric constant. $\mathbf{u}$ is the fluid velocity. $F$, $R$, $T$, $\mu$, and $p$ are the Faraday's constant, gas constant, temperature, liquid viscosity, and pressure, respectively. $\mathbf{J}_i$, $C_i$, $D_i$, and $z_i$ are the ionic flux, concentration, diffusion coefficient, and valence of ionic species $i$ ($K^+$ and $Cl^-$ ions), respectively.

The total current $I$ under different voltages $V$ was obtained through the integration of ionic flux from cations and anions over the boundary of the reservoir with equation 5.

$$I = \int_S F \left( \sum_i^2 z_i \mathbf{J}_i \right) \cdot \mathbf{n} \, dS \quad (5)$$

where $S$ represents the reservoir boundary and $\mathbf{n}$ is the unit normal vector.

To figure out the modulation of ion transport from charged exterior surfaces, eight different simulation models have been built with differently charged surfaces (Figure 1).[24, 25, 34] Referring to the direction of the electric field applied across the nanopore, three nanopore walls are named the inner-pore surface, the exterior surface on the voltage side (surface$_V$), and the ground side (surface$_G$). Then the eight simulation models were constructed according to the charged status of individual surfaces, i.e. nanopores with uniformly charged surfaces (ACS), none charged surfaces (NCS), only charged inner-pore surface (ICS), both charged exterior surfaces (ECS), only charged exterior surface on the voltage side (ECS$_V$), only charged exterior surface on the ground side (ECS$_G$), both charged inner-pore surface and exterior surface on the ground side (IECS$_G$), and both charged inner-pore surface and exterior surface on the voltage side (IECS$_V$). The surface charge density was set as −0.08 C/m$^2$.[35-37] The mesh strategy and boundary conditions used in the simulations were the same as in previous studies (Figure S1 and Table S1).[24, 25, 34] For charged surfaces of the nanopore, a mesh size of 0.1 nm was applied. The total mesh nodes exceeded 1,000,000.

## III. RESULTS AND DISCUSSION

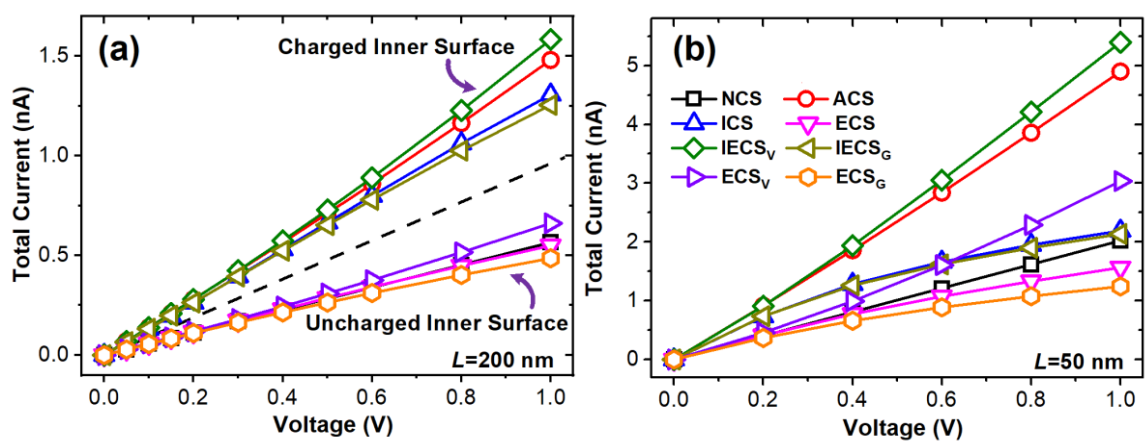

Figure 2 Current-voltage (I-V) curves in the eight simulation models with different charged surfaces. (a-b) Total current through nanopores with 200 (a) and 50 nm (b) in length.

Taking advantage of the simulation method, it is convenient to obtain the ion current contributed by cations, anions, or both, which corresponds to the $K^+$ ion current, $Cl^-$ ion current, or total current in this work. Figure 2 shows the current-voltage (I-V) curves obtained in nanopores with different charged surfaces under pore lengths of 200 and 50 nm. I-V curves through nanopores with other lengths (20, 100, 500, and 1000 nm) are plotted in Figure S2. From Figure 2a, in the 200-nm-long nanopores, eight I-V curves can be divided into two groups depending on the charged status of the inner-pore wall.[23] In each group, the individual I-V curve shares almost equal values, especially under low voltages. The same trend is also found in nanopores longer than 200 nm (Figure S2). In these long nanopores, the ion current is mainly determined by surface charges on the inner pore surface. Compared to the nanopore with uniformly neutral surfaces (NCS case), the ICS case with a surface charge density of $-0.08$ C/m$^2$ can facilitate ion transport significantly. Because ion concentration inside the charged pore (Figure S3) is dramatically enhanced which can induce strong surface conductance along the charged inner-pore wall, ion current is promoted by ~1.3 times that in the NCS case. For long nanopores, the charged exterior surfaces have no apparent modulation to ion current. This is why the charged exterior surface was seldom considered in earlier simulations and theoretical analyses.[8, 12-14, 17, 18]

Nanopores with sub-100 nm in length are commonly required in biosensing thanks to their higher sensitivity,[35, 38] where the applied voltage across the nanopores can reach as high as 2 V. In our previous research on osmotic energy conversion under salt gradients with nanopores shorter than 100 nm,[25, 32, 34] the charged exterior surface of the low-concentration side can promote the diffusion of counterions effectively, which improves both the output power and efficiency of the energy conversion. Here, under electric fields, with the pore length shrinking to 100 nm or less (Figure S2 and Figure 2b), the influence of exterior surface charges on ion transport through nanopores becomes more considerable. From the I-V curves through 50-nm-long nanopores, the ion current caused by the charged exterior surface on the voltage side ($ECS_V$ case) is greater than that in the case with only a charged inner surface (ICS case) at 0.6 V and above.

Following the same strategy in Figure 2a, the behaviors of ion current obtained from the eight cases with 50 nm in length are analyzed separately based on the charged status of inner-pore walls. For nanopores with neutral inner surfaces, the individual charged exterior surface on the voltage side ($ECS_V$ case) and ground side ($ECS_G$ case) can enhance and suppress the ion current by ~50% and ~38.5% compared to that in the neutral nanopore (NCS case) at 1 V (Figure S4), respectively. For the ECS case with both charged exterior surfaces which can be treated as the combination of $ECS_V$ and $ECS_G$ cases, the obtained ion current is inhibited compared to that of the NCS case. This indicates that in sub-100 nm long nanopores with neutral inner surfaces, the suppression caused by the charged exterior surface at the ground side on ion transport is stronger than the promotion by the charged exterior surface at the voltage side.

When the inner surface of the nanopore is charged, the induced limiting current of the I-V curve by ion concentration polarization in the ICS case[39] is smaller than the enhanced current in the case of $ECS_V$. After the exterior surface is also charged, from Figure 2b, the charged surface$_V$ (IECS$_V$ case) can induce a strong enhancement on the ion current that overcomes the limiting phenomenon. The inhomogeneity in charged pore surfaces causes more counterions to enter than exit the nanopore. The total current of the IECS$_V$ case is the largest among all cases, which is ~145% higher than that of the ICS case at 1 V. However, there is no obvious modulation to the current from charged exterior surfaces at the ground side. The current of the IECS$_G$ case is almost equal to that of the ICS case. As a result of the weak suppression effect on ion current by the charged exterior surface on the ground side, the ion current in the ACS case is slightly smaller than that of the IECS$_V$ model.

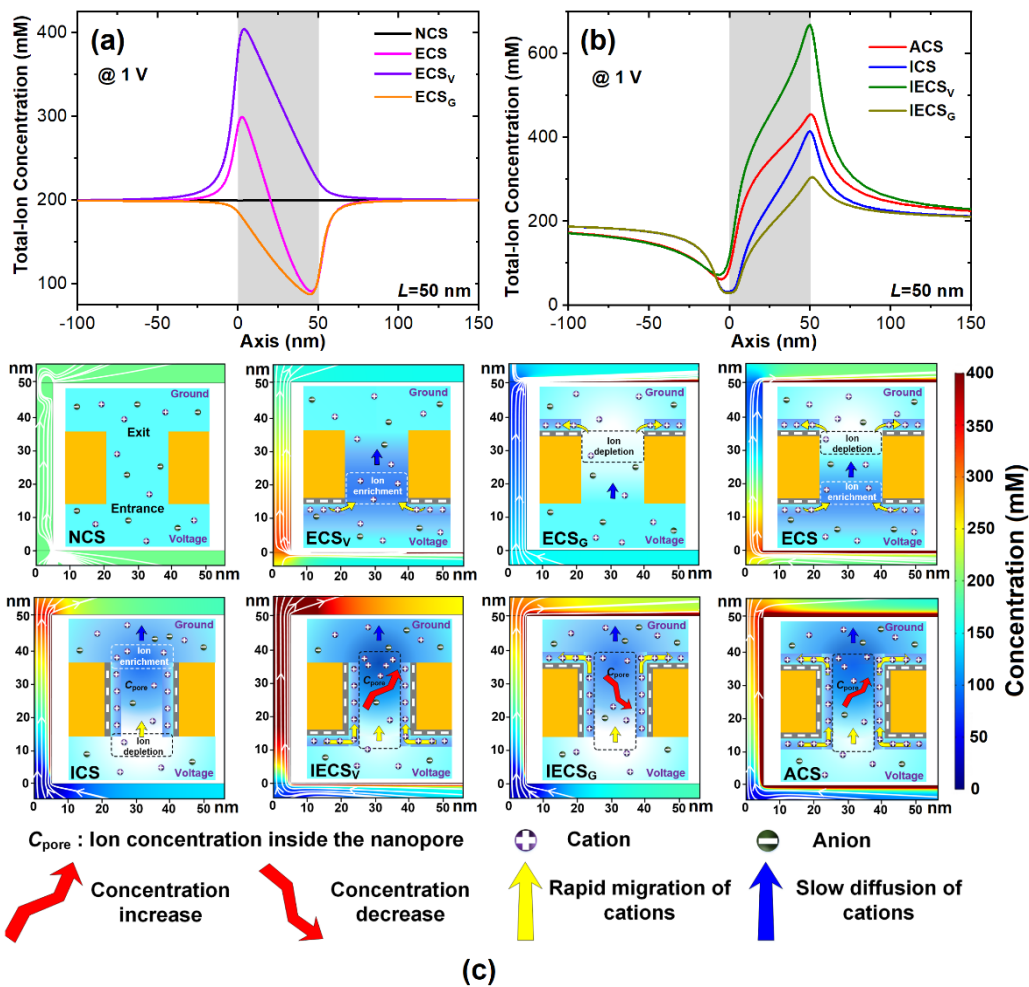

Figure 3 Total-ion concentration along the pore axis in nanopores w/ and w/o charged inner surface at 1V and illustrations of the ion transport in eight cases. (a) Nanopores with neutral inner-pore walls. (b) Nanopores with charged

inner-pore walls. Pore regions are shown in light gray. The pore length and diameter are 50 and 10 nm, respectively. The solution is 100 mM KCl. (c) Illustrations of the ion transport and 2D concentration distributions of total ions (color maps) and streamline (white lines) in eight cases, in which the $C_{pore}$ represents the ion concentration inside the nanopore. The light and dark blue regions show ion depletion and enrichment, respectively. The yellow and blue arrows represent the rapid migration and slow diffusion of cations, respectively. The red arrows only represent the increase and decrease in ion concentration in the nanopore compared to the ICS case. The locations where counterions enter and exit the nanopore are defined as the entrance and exit of the nanopore.

Figures 3a and 3b show the total-ion concentration along the pore axis in nanopores w/ and w/o charged inner surface in eight cases, respectively. The modulation mechanism of the individual charged surfaces on ion transport is shown in Figure 3c. From Figure 3a, the neutral nanopore (NCS) has no ion selectivity, and the ion concentration inside the nanopore is the same as the bulk concentration. For the $ECS_V$ case, EDLs near charged exterior surfaces on the entrance side act as an ion pool. A large number of counterions inside the EDLs can be transported to the pore entrance along the charged exterior surface (Figure S5). The relatively weak ionic movement from the inside to the outside of the pore induces ion enrichment at the front part of the pore (Figure 3c, $ECS_V$). While the EDL region on the exit side of the $ECS_G$ case provides a fast passageway for the flowing out of counterions.[26] The strong migration of counterions along the charged exterior surface results in apparent ion depletion at the back part of the pore (Figure 3c, $ECS_G$). The ion concentration distribution inside the ECS pore demonstrates that the modulation of ion transport is influenced simultaneously by each individual charged surface. In Figure 3a, both the ion enrichment and depletion appear in the nanopore of the ECS case. The intersection of the axis concentration and bulk values (200 mM) of the ECS case is located near the pore entrance which implies the more obvious ion depletion effect induced by the charged exterior surface at the ground side. Because ion depletion happens near the pore exit which can cause a relatively larger electric field strength (Figure

S6a), the accelerated ionic migration from the pore exit to the bulk tends to aggravate the ion depletion. This positive feedback induces the dominant modulation in ion current from the charged exterior surface on the ground side.

From Figure 3b, owing to the high ionic selectivity of nanopores to counterions caused by charged inner-pore walls, obvious ICP happens which shows the ion depletion and enrichment at the pore entrance and exit, respectively. For the cases with combined charged inner and exterior surfaces, in addition to ICP, ion enrichment and depletion inside nanopores should also be considered. For the $IECS_V$ case, the ion enrichment zone is not formed at the nanopore entrance like in the $ECS_V$ case. Instead, the ion concentration gets depleted and enriched at the entrance and exit of the nanopore, respectively, which is similar to the ICS case (Figure 3c, $IECS_V$). However, the exterior surface conductance induced by the EDLs near the charged exterior surface can promote tangential ionic migration along the exterior surface. Large amounts of counterions are replenished at the pore entrance which induces weaker ion depletion than that in the ICS case. Due to the strong ion transport along the charged inner surface, the ion concentration increases rapidly inside the nanopore, which causes more obvious ion enrichment at the exit side with a peak value of ~1.6 times that in the ICS case. In the case of $IECS_G$, the electric field strength at the entrance of the nanopore is stronger than that in the case of $IECS_V$ (Figure S6b). The stronger electric field strength provides counterions with a larger migration velocity, which results in a lower ion concentration at the entrance of the $IECS_G$ pore than that of the $IECS_V$ case. In addition, the enhanced ion transport along the charged exterior surface results in a drop of ~26% in the concentration enrichment at the pore exit compared to that in the ICS case. Because the dominant electrical resistance appears at the nanopore entrance with the lowest ion concentration, the current from both $IECS_G$ and ICS cases shares almost the same values, which are much smaller than those in the ACS and $IECS_V$ models (Figure 2b). For the ACS case with uniformly charged pore walls, the ion concentration distribution shares a similar trend with that in the $IECS_V$ case.[25] The additional charged surface on the ground side promotes the exit of counterions, which weakens the ion enrichment at the exit side. The above results show that in nanopores with charged inner surfaces, the suppression from the charged surface$_G$ on the ion current is weakened, while the promotion from the charged surface$_V$ on the ion current is improved.

Here, our results agree well with the synergistic effect of both charged inner surface and exterior surface found by Li et al.[28] The presence of charged exterior surfaces significantly promotes the ion transport to enter and exit the nanopore. In cases with neutral inner-pore walls, due to the lack of EDLs inside nanopores, counterions cannot be transported effectively which induces ion enrichment at the pore entrance, ion depletion at the pore exit, or both. In systems with charged inner-pore surfaces, EDLs inside nanopores not only increase the intro-pore ion concentration but also provide a fast passageway for the transport of counterions. For nanopores with both charged inner-pore walls and exterior surfaces, the influences from charged inner-pore walls and exterior surfaces can be additive. Continuous EDLs near charged surfaces enhance the ion transport through the nanopore significantly.

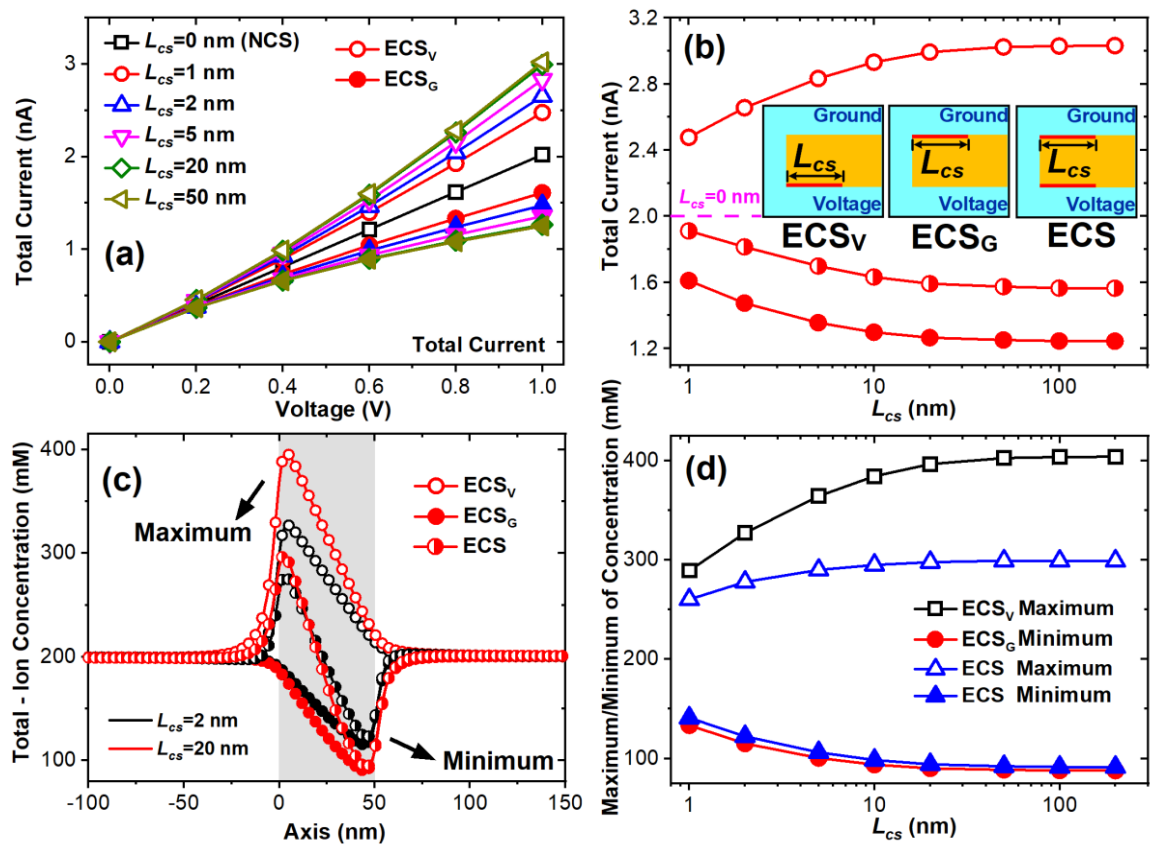

Figure 4 Ionic behaviors in the three cases with charged exterior surfaces and neutral inner-pore surfaces. Widths ($L_{cs}$) of charged ring regions near

nanopores on the exterior surfaces are varied from 0 to 50 nm. (a) I-V curves obtained from $ECS_V$ and $ECS_G$ models. (b) Total ion current at 1 V from the three models with various $L_{cs}$. The inset shows the scheme of three simulation models with neutral inner-pore walls. Charged surfaces are shown in red. (c) Total ion concentration distribution along the pore axis at 1 V in the three models. (d) Maximum and minimum concentration distributions from three simulation cases with various $L_{cs}$. The pore length and diameter are 50 and 10 nm, respectively. The solution is 100 mM

Our simulation results show that besides inner-pore walls the charged exterior surfaces can provide significant influences on ion transport in short nanopores, which depend on the charged status of the inner-pore surfaces. Based on the wide applications of nanopores in biosensing,[40, 41] energy conversion,[42] and related fields,[43, 44] porous membranes are the common nanofluidic platforms to reach high throughputs. Due to the valuable area of porous membranes, it is particularly important to reasonably arrange the distribution of nanopores. Here, we have investigated influences from the width ($L_{cs}$) of charged ring regions on exterior surfaces on the ionic behaviors through nanopores to find its effective value.[24, 25]

As shown in Figure 4a, I-V curves are obtained from two cases with neutral inner-pore walls but differently charged exterior surfaces with various widths. With charged exterior surfaces on the voltage and ground sides, the ion current increases and decreases with the enlargement of $L_{cs}$, respectively. As the applied voltage enhances, both the current increase and decrease become more apparent, which saturate at $L_{cs}$~20 nm at 1 V (Figure 4b). As the $L_{cs}$ reaches 20 nm, the ion current from the $ECS_V$ and $ECS_G$ cases increases and decreases by ~48.0% and ~37.6%, respectively. For the ECS case with both charged exterior walls, the current difference from that in the NCS case increases gradually with $L_{cs}$. Because of the dominant effect of the exterior surface on the ground side, ion current also reaches its saturation at $L_{cs}$ ~20 nm, with ~21.3 % in decrease.

From Figure S7, through analyzing individual I-V curves contributed by $K^+$ and $Cl^-$ ions, the charged exterior surfaces on the voltage and ground sides can promote and inhibit the transport of anions and cations simultaneously. With the increase of the $L_{cs}$, both promotion and inhibition from charged exterior surfaces to ion transport becomes more significant, corresponding to the modulation in ion concentration as shown in Figure 4c and 4d. With the $L_{cs}$ increasing from 0 to 20 nm, ion enrichment and depletion inside the pore are promoted from 200 to ~390 mM in the $ECS_V$ case, and suppressed from 200 to ~90 mM in the $ECS_G$ case, respectively. In the ECS models, the minimum concentration values share a similar trend to those in the $ECS_G$ case. While, based on the combined effects from both exterior surfaces, ion enrichment is weakened which saturates at $L_{cs}$ ~ 10 nm.

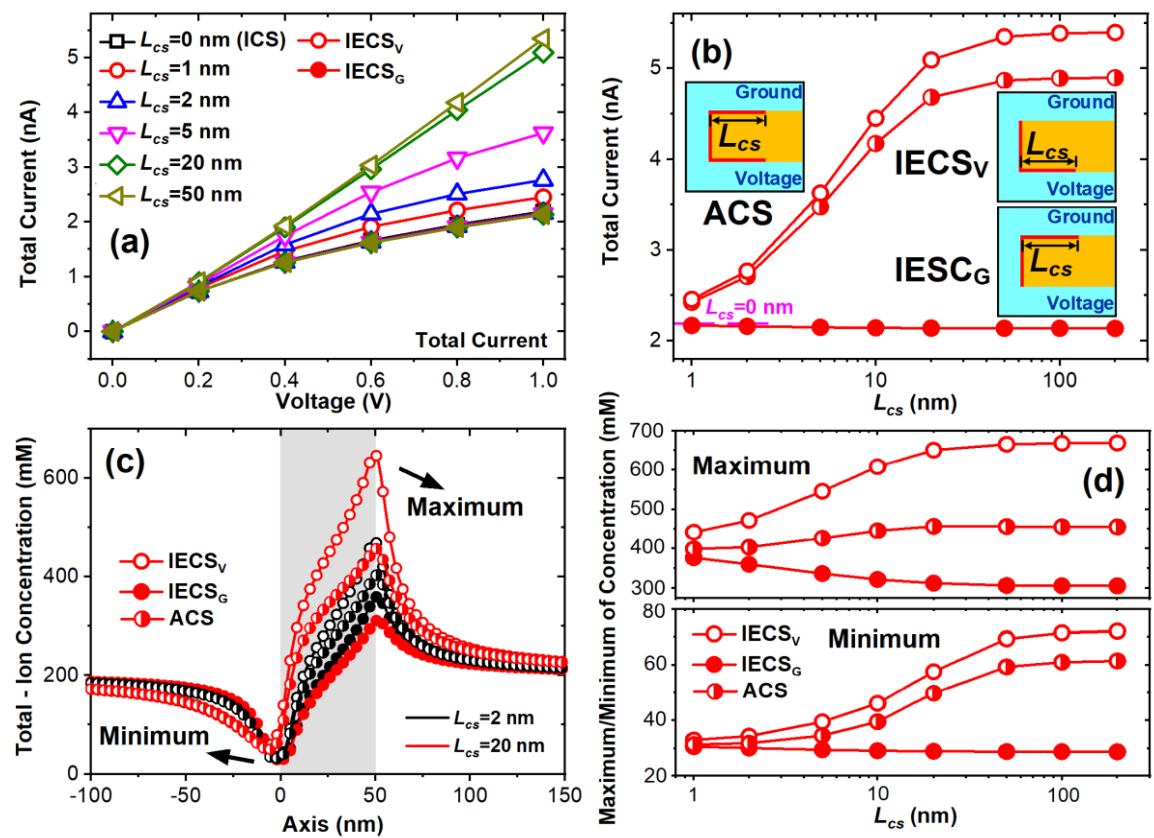

Figure 5. Ionic behaviors in the three cases with charged exterior surfaces and inner-pore surfaces. Widths ($L_{cs}$) of charged ring regions near nanopores on the exterior surfaces are varied from 0 to 50 nm. (a) I-V curves obtained from $IECS_V$ and $IECS_G$ models. (b) Total ion current at 1 V from the three models with various $L_{cs}$. The inset shows the scheme of three simulation models with neutral inner-pore walls. Charged surfaces are shown in red. (c) Total ion concentration distribution along the pore axis at 1 V in the three

models. (d) Maximum and minimum concentration distributions from three simulation cases with various $L_{cs}$. The pore length and diameter are 50 and 10 nm, respectively. The solution is 100 mM KCl.

Following the same strategy, the ionic behaviors have also been investigated through nanopores with charged inner-pore walls and charged exterior surfaces (Figure 5). When the inner-pore surface is charged, the dominant role in current modulation is from the charged exterior surface on the voltage side. In Figure 5a, as the $L_{cs}$ increases almost constant ion current is obtained in the IECS$_G$ case, which implies that the charged exterior surface on the ground side has almost negligible modulation on the current. As a result, the ion current of the ACS case has the same trend as that in the IECS$_V$ case (Figure S8). From Figure 5b, the total current in all three cases gets saturated at $L_{cs}$ ~20 nm, where the current from the IECS$_V$ and ACS cases is increased by ~132.4% and ~113.7%, respectively, and that from IECS$_G$ case is decreased by ~2.4%.

Corresponding ion concentration distributions under various $L_{cs}$ are plotted in Figures 6c and 6d. Due to the strong migration of ions along charged exterior surfaces (Figure S5), with the increase of $L_{cs}$, the charged exterior surface on the voltage side weakens the ion depletion at the pore entrance and enhances the ion enrichment inside the pore effectively, which induces significate promotion in ion current. While the exterior surface on the ground side mainly decreases the ion enrichment at the pore exit. Based on the dependence of ion current on the ion distribution, the dominant regions of the electric resistance in the IECS$_G$ and IECS$_V$ cases are the depletion at the pore entrance and intra-pore enrichment, respectively. In all three cases, the concentration distribution almost reaches its equilibrium at $L_{cs}$ ~20 nm.

## IV. CONCLUSIONS

Benefiting from the finite element method, we systematically studied the influence of charged external surfaces on ion transport with different simulation models. As the pore length decreases from 200 to 50 nm, charged exterior surfaces play a considerable role in the modulation of ion current. We find that charged exterior surfaces exhibit different modulations of ion transport depending on the charged status of inner-pore walls. Due to the fast ion transport in EDLs along charged surfaces, for nanopores with neutral inner-pore walls, the charged exterior surface on the voltage and ground sides can cause obvious ion enrichment and depletion inside nanopores which promotes and inhibits the ion conductance. Charged surface$_G$ has a stronger inhibitory effect on the total current. Because charged inner-pore walls induce enhanced ion transport in the axial direction, ICP occurs at the nanopore orifice. Modulation mechanisms from charged exterior surfaces changed in nanopores with charged inner surfaces. The charged exterior surface on the voltage side promotes ion conductance significantly by simultaneously weakening the ion depletion at the pore entrance and enhancing the enrichment inside the nanopore. However, the influence from the charged exterior surface on the ground side is negligible. By adjusting the area of the charged exterior surfaces, the ion current and ion concentration inside nanopores first increase with the $L_{cs}$ and then reach equilibrium at the effective charged widths of ~20 nm, which shows weak dependence on the charged status of inner-pore surfaces. Our results may help in engineering nanofluidic devices for biosensing, desalination, and other related fields.

## SUPPLEMENTARY MATERIAL

See supplementary material for simulation details and additional simulation results.

## ACKNOWLEDGMENT

This work was supported by the National Natural Science Foundation of China (52105579), the Natural Science Foundation of Shandong Province (ZR2020QE188), the Guangdong Basic and Applied Basic Research Foundation (2023A1515012931), the Qilu Talented Young Scholar Program of Shandong University, and the open Foundation of Fujian Key Laboratory of Functional Marine Sensing Materials, Minjiang University (MJUKF-FMSM202205).

## AUTHOR DECLARATIONS

## CONFLICT OF INTEREST

The author has no conflicts to disclose.

## AUTHOR CONTRIBUTIONS

**Long Ma:** Simulation, Investigation, Data curation, Writing - original draft, review & editing. **Zhe Liu:** Data curation. **Bowen Ai:** Data curation. **Jia Man:** Investigation. **Jianyong Li:** Investigation. **Kechen Wu:** Investigation. **Yinghua Qiu:** Conceptualization, Methodology, Resources, Writing - original draft, review & editing, Supervision, Funding acquisition.

## DATA AVAILABILITY

The data that support the findings of this study are available from the corresponding author upon reasonable request.